\documentclass[prl,twocolumn,showpacs,superscriptaddress,preprintnumbers,floatfix,amsmath,amssymb]{revtex4-1}

\usepackage{amssymb}
\usepackage{graphicx}
\usepackage{dcolumn}
\usepackage{bm}
\usepackage{color}
\hfuzz=\maxdimen
\begin{document}

\title{Evidence of Spontaneous Vortex Ground State in An Iron-Based Ferromagnetic Superconductor}

\author{Wen-He Jiao}
\affiliation{Department of Physics, Zhejiang University, Hangzhou
310027, P. R. China} \affiliation{Department of Physics, Zhejiang University of Science and Technology, Hangzhou 310023, P. R. China}

\author{Qian Tao}
\affiliation{Department of Physics, Zhejiang University, Hangzhou 310027, P. R. China}

\author{Zhi Ren}
\affiliation{Institute for Natural Sciences, Westlake Institute for Advanced Study, Hangzhou 310013, P. R. China}
\author{Yi Liu}
\affiliation{Department of Physics, Zhejiang University, Hangzhou 310027, P. R. China}

\author{Guang-Han Cao}
\email[]{ghcao@zju.edu.cn}
\affiliation{Department of Physics, Zhejiang University, Hangzhou
310027, P. R. China}
\affiliation{State Key Lab of Silicon Materials, Zhejiang University, Hangzhou 310027, P. R. China}
\affiliation{Collaborative Innovation Centre of Advanced Microstructures, Nanjing 210093, P. R. China}

\date{\today}

\begin{abstract}
\textbf{Spontaneous vortex phase (SVP) is an exotic quantum matter in which quantized superconducting vortices form in the absence of external magnetic field. Although being predicted theoretically nearly 40 years ago, its rigorous experimental verification still appears to be lacking. Here we present low-field magnetic measurements on single crystals of the iron-based ferromagnetic superconductor Eu(Fe$_{0.91}$Rh$_{0.09}$)$_{2}$As$_{2}$
which undergoes a superconducting transition at $T_\mathrm{sc}$ = 19.6 K followed by a magnetic transition at $T_\mathrm{m}$ = 16.8 K. We observe a characteristic first-order transition from a Meissner state within $T_\mathrm{m}<T<T_\mathrm{sc}$ to an SVP below $T_\mathrm{m}$, under a magnetic field approaching zero. Additional isothermal magnetization and ac magnetization measurements at $T\ll T_\mathrm{sc}$ confirm that the system is intrinsically in a spontaneous-vortex ground state. The unambiguous demonstration of SVP in the title material lays a solid foundation for future imaging and spectroscopic studies on this intriguing quantum matter.}

\end{abstract}

\pacs{74.70.Xa, 74.25.Ha, 75.30.-m}


\maketitle

Spontaneous vortex (SV) phase, originally predicted by theoretical investigations~\cite{Varma1,Tachiki,Varma2,Kuper}, is an exotic quantum matter in which superconducting vortices form in the absence of external magnetic field, which can be qualitatively different from those induced by an external field~\cite{svglass}. While SV state was also predicted to be present in the pseudogap phase of cuprates due to local spins of the paramagnetic phase~\cite{wzy}, self-induced vortices are mostly generated by an internal magnetic field, $H_{\mathrm{int}}$ = 4$\pi M$, due to the spontaneous magnetization $M$. This means that the prerequisite of realization of an SV state is that superconductivity (SC) coexists with magnetic order, the latter of which at least gives rise to a \emph{ferromagnetic} component. Such a coexistence is rare because of the antagonism between SC and ferromagnetism (FM). Additional requirements for observation of the SV phenomenon include yet, are not limited to, (1) the SC alone (a hypothetical nonmagnetic analog without any internal field) belongs to the second type with intrinsic lower and upper critical fields ($H_{\mathrm{c1}}^*$ and $H_{\mathrm{c2}}^*$); and (2) the internal magnetic field strength lies in the range of $H_{\mathrm{c1}}^* < H_{\mathrm{int}} <  H_{\mathrm{c2}}^*$.

\begin{figure}
\includegraphics[width=8cm]{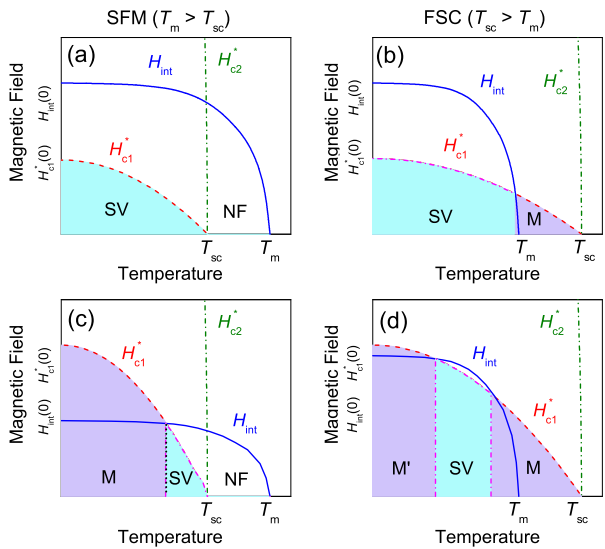}
\caption{\textbf{Classification of magnetic superconductors that may host the spontaneous vortex (SV) phase.} The left panels show superconducting ferromagnets (SFMs), while the right panels describe ferromagnetic superconductors (FSCs). (a) and (b), $H_{\mathrm{int}}(0) > H_{\mathrm{c1}}^*(0)$; (c) and (d), $H_{\mathrm{c1}}^*(0) > H_{\mathrm{int}}(0)$. See the text for the definitions of $H_{\mathrm{int}}$ and $H_{\mathrm{c1}}^*$. M and M' denote Meissner states, and NF stands for a non-superconducting ferromagnetic state. Here the upper critical fields, $H_{\mathrm{c2}}^*(0)$, are assumed to be much larger than $H_{\mathrm{c1}}^*(0)$ and $H_{\mathrm{int}}(0)$.}
\label{fig1}
\end{figure}

Materials that bear both SC and FM generally have distinct superconducting critical temperature $T_\mathrm{sc}$ and magnetic transition temperature $T_\mathrm{m}$. If $T_\mathrm{sc} > T_\mathrm{m}$, they are traditionally called ``ferromagnetic superconductors" (FSCs)~\cite{Buzdin85}. Otherwise, the terminology ``superconducting ferromagnets" (SFMs) are often employed \cite{Felner}. According to this classification and, with the consideration of the relative strength between $H_{\mathrm{c1}}^*$ and $H_{\mathrm{int}}$, possible existence of an SV phase is schematically depicted in Fig.~\ref{fig1} for an extremely type-II superconductor in which $H_{\mathrm{c2}}^*(0) \gg H_{\mathrm{c1}}^*(0)$ holds. In the cases of $H_{\mathrm{int}}(0) > H_{\mathrm{c1}}^*(0)$, as shown in the panels (a) and (b), the SV phase can be realized as a ground state. If $H_{\mathrm{int}}(0) < H_{\mathrm{c1}}^*(0)$, however, possible SV phase appears only at finite temperatures above zero, as is seen in the panels (c) and (d).

There have been a few systems that may host an SV state. In the SFMs, Ru-containing cuprates~\cite{Felner,Bernhard} and U-based UCoGe~\cite{UCoGe-sv,UCoGe2012}, which can be categorized into scenarios (c) and (a) in Fig.~\ref{fig1}, respectively, were argued to have an SV state on the basis of magnetic measurements. For FSCs, however, evidence of SV phase from bulk magnetic measurements is still lacking, although theoretical~\cite{Varma3} and experimental~\cite{ErNi2B2C-sv} investigations suggested an SV state in the weakly ferromagnetic superconductor ErNi$_{2}$B$_{2}$C. In fact, rigorous demonstration of a bulk SV phase by magnetic measurements is challenging primarily because an external field, which by itself induces vortices and, possibly changes the magnetic state, has to be applied. In general, one needs to demonstrate the existence of SV state as \emph{the external field approaches zero}, which requires a sufficiently high measurement precision. This issue becomes more stringent in the cases above where the internal field generated by the small ferromagnetic component is very weak (e.g., the spontaneous magnetization of UCoGe is $\sim$0.04 $\mu_\mathrm{B}$/U, corresponding to $\sim$30 Oe field). Furthermore, the magnetic measurements always encounter the interferences of ferromagnetic domains~\cite{UCoGe2012}.

As was first pointed out by Ng and Varma~\cite{Varma3}, nevertheless, the SV phase can be manifested by the unique first-order phase transition from a Meissner state to an SV phase, which can be possibly seen in cases of Fig.~\ref{fig1}(b-d). The first-order transition is expected to accompany with a thermal hysteresis that may be easily captured experimentally. Indeed, a thermal hysteresis in magnetic susceptibility was observed in the SFM RuSr$_{2}$GdCu$_{2}$O$_{8}$, which is interpreted as a characteristic of SV state~\cite{Bernhard}. However, the observed phenomenon was sample dependent and, the polycrystalline samples employed expose the flaw: the magnetic-flux pinning by grain boundaries might also account for the phenomenon~\cite{Papageorgiou}. Therefore, detection of a Meissner-to-SV transition should be done at least using single crystalline samples.

In this context, the recently discovered FSCs in doped EuFe$_{2}$As$_{2}$ systems~\cite{Dressel}, which show a remarkable coexistence of SC and strong FM in a broad temperature range (note that the temperature window for probing an SV phase is mostly below 2.5 K in previous systems~\cite{UCoGe-sv,UCoGe2012,ErNi2B2C-sv}), provide a desirable platform to look into the SV state. Through either P doping at As site~\cite{Eu122P-rz} or transition-metal (such as Ru, Co, Rh, and Ir) doping at Fe site~\cite{Eu122Co-js,Eu122Ru, Eu122Rh,Eu122Ir-jwh,Eu122Ir-hossian}, SC can be induced with a $T_{\rm sc}$ between 20 and 30 K, and the Eu$^{2+}$ spins (with $S$ = 7/2) become ferromagnetically ordered at $T_{\rm m}$ a few kelvins lower. Although there were debates on details of the magnetic order~\cite{Dressel,Felner2011,terashima,jeevan,tokiwa,zapf}, recent x-ray resonant magnetic scattering and neutron diffraction studies \cite{Eu122Co-jwt,Eu122P-xms,Eu122P-neutron,Eu122Ir-jwt} show that the Eu$^{2+}$ spins always align ferromagnetically along the $c$ axis with an ordered moment of about 7 $\mu_{\rm B}$. The ferromagnetic ordering gives rise to a large spontaneous magnetization that generates an internal field of $H_{\mathrm{int}} \approx$ 9,000 Oe along the $c$ axis, well above the expected $H_{\mathrm{c1}}^{*}(0)$ of $\sim$150 Oe \cite{Hc1}. Additional important advantage of the iron-based FSC is that the high-quality single crystals are easily accessible~\cite{Eu122Co-js,Eu122Ru, Eu122Rh,Eu122Ir-jwh}. Note that the internal-field direction induces superconducting vortices within the FeAs layers. As such, the magnetic measurements can be limited to those under external fields parallel to the $c$ axis, which greatly simplifies the interpretation of the measurement result.
\\

\textbf{Results}

\begin{figure}
\includegraphics[width=8cm]{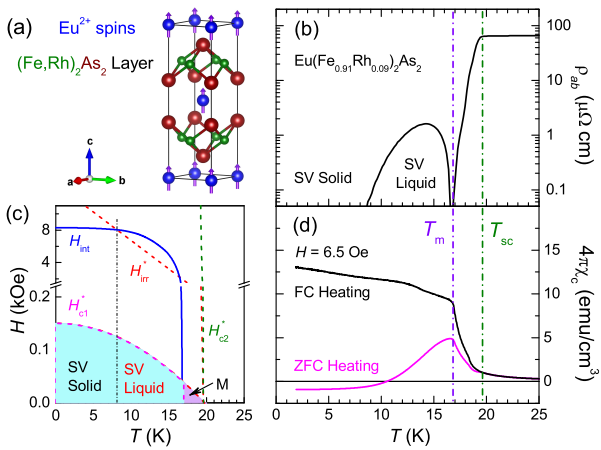}
\caption{\textbf{Characteristic of the ferromagnetic superconductor Eu(Fe$_{0.91}$Rh$_{0.09}$)$_{2}$As$_{2}$ in relation with a spontaneous vortex phase.} (a) The crystal and magnetic structure. (b) The superconducting resistive transition at $T_{\rm sc}$, followed by a resistivity revival below $T_{\rm m}$ (note the logarithmic scale for resistivity). (c) Schematic $H-T$ diagram showing the internal field $H_{\mathrm{int}}$ (solid blue line), in comparison with the hypothetical irreversible field $H_{\mathrm{irr}}^*$ (assuming $H_{\mathrm{int}}$ = 0) as well as the hypothetical lower and upper critical fields, $H_{\mathrm{c1}}^*$ and $H_{\mathrm{c2}}^*$. SV and M denote the spontaneous-vortex phase and the Meissner state, respectively. (d) Temperature dependence of the dc magnetic susceptibility measured while heating up, with both field-cooled (FC) and zero-field-cooled (ZFC) histories. The demagnetization effect has been taken into account.}
\label{fig2}
\end{figure}

We employed an optimally Rh-doped single crystal of Eu(Fe$_{0.91}$Rh$_{0.09}$)$_{2}$As$_{2}$ with $T_{\mathrm{sc}}$ = 19.6 K, $T_{\mathrm{m}}$ = 16.8 K, and a saturation magnetization $M_{\mathrm{sat}} = 6.5$ $\mu_{\mathrm{B}}/\mathrm{Eu}$ ~\cite{Eu122Rh}. The saturation magnetization is close to $gS$ = 7.0, which tells that the Eu$^{2+}$ spins align ferromagnetically, similar to other Eu-containing FSC ~\cite{Eu122Co-jwt,Eu122P-xms,Eu122P-neutron,Eu122Ir-jwt}, as shown in Fig.~\ref{fig2}(a). The superconducting transition of in-plane resistivity is plotted in Fig.~\ref{fig2}(b). The relatively broadened resistive transition seems to be related to the Eu-spin exchange field which suppresses the $T_{\mathrm{sc}}$ value (note that $T_{\mathrm{sc}}$ is 21.9 K for the optimally Rh-doped SrFe$_{2}$As$_{2}$~\cite{whh}). Below $T_{\mathrm{m}}$, the ferromagnetic ordering leads to a re-appearance of resistivity [Fig.~\ref{fig2}(b)]. The maximum of the reentrant resistivity is only 1/40 the normal-state value, indicating that it is by no means a recovery of the normal state, instead, it is associated with the SV formation. Specifically speaking, the revival of resistivity comes from the vortex flow in an \emph{SV liquid} state. With decreasing temperature, $H_{\mathrm{irr}}^*$ surpasses $H_{\mathrm{int}}$, as shown in Fig.~\ref{fig2}(c), making the vortices frozen, hence zero-resistance state is achieved below $\sim$8 K. Note that the SV scenario naturally explains various resistivity states below $T_{\mathrm{m}}$~\cite{Dressel,Eu122P-rz,Eu122Co-js,Eu122Ru, Eu122Rh,Eu122Ir-jwh,Eu122Ir-hossian}, some of which show absence of the resistivity reentrance~\cite{Eu122Ir-jwh,terashima,jeevan,tokiwa}, depending on the doping levels and physical pressures. As is seen, the absence of reentrant behaviour is more easily to be observed in P-doped EuFe$_{2}$As$_{2}$~\cite{terashima,jeevan,tokiwa} where $T_{\mathrm{sc}}$ is significantly higher than $T_{\mathrm{m}}$ such that $H_{\mathrm{irr}}^*>H_{\mathrm{int}}$ is satisfied.

The dc magnetic susceptibility shows a kink for the field-cooling (FC) protocol and a peak for the zero-field-cooling (ZFC) protocol at $T_{\rm m}$, as shown in Fig.~\ref{fig2}(d). This can be interpreted as the formation of antiparallel ferromagnetic domains~\cite{Eu122Ir-jwh}. Because of the proximity between SC and FM, the superconducting transition is not distinctly seen in the dc magnetic measurements (although it was directly observable at very low fields \cite{Eu122Ir-jwh}). Nevertheless, $\chi_c^{\mathrm{FC}}$ and $\chi_c^{\mathrm{ZFC}}$ bifurcate just at $T_{\rm sc}$, owing to the magnetic-flux pinning effect. The superconducting magnetic shielding effect below $T_{\rm sc}$ is confirmed by the following ac susceptibility measurement.

Since the internal field generated by the Eu$^{2+}$-spin FM is much stronger than the expected $H_{\mathrm{c1}}^{*}(0)$, as described above, the SV state is stabilized once the FM develops. On the other hand, the internal field vanishes for $T > T_{\rm m}$, hence it is in a Meissner state at zero external field in the temperature range $T_{\rm m} < T < T_{\rm sc}$, as shown in Fig.~\ref{fig2}(c). Therefore, a transition from the Meissner state to the SV phase definitely occurs as temperature decreases. During the transition, the spontaneous vortices (SVs) suddenly penetrate the crystal's interior, which gives rise to a unique first-order transition with a magnetization discontinuity at around $T_{\rm m}$ for the ideal case with single magnetic domain. In the case of a large sample with multi-domains, nevertheless, a ``continuous" change with an obvious thermal hysteresis is expected because of the latent heat in the first-order transition. The possible thermal hysteresis from the domain-wall depinning can be avoided by employing magnetic fields that are much lower than the coercive field ($\sim$200 Oe~\cite{Eu122Ir-jwh}).

\begin{figure}
\includegraphics[width=8cm]{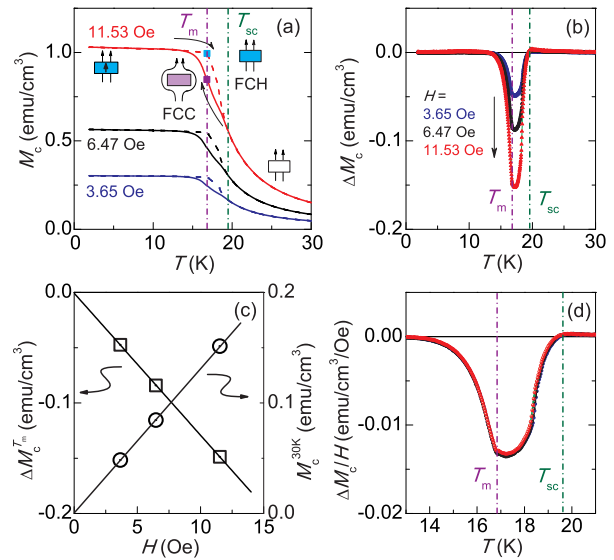}
\caption{\textbf{Evidence of the first-order transition from a Meissner state to a spontaneous vortex phase with decreasing temperature in Eu(Fe$_{0.91}$Rh$_{0.09}$)$_{2}$As$_{2}$ crystals.} (a) Field-cooling magnetization ($M_c$) on both heating (FCH) and cooling (FCC) processes under magnetic fields along the $c$ axis. The rectangles with arrows schematically represent different statuses of sample in the presence of external field (multi-domains and pinned fluxes are not shown). See the text for the inserted cartoon pictures (b) The magnetization difference $\Delta M_{c}$ between the FCH and FCC data. (c) $\Delta M_{c}$ at $T_\mathrm{m}$ (left axis) and $M_c$ at 30 K (right axis) as functions of the applied field. (d) $\Delta M_{c}/H$ versus $T$ in an expanded temperature range.}
\label{fig3}
\end{figure}

As shown in Fig.~\ref{fig3}(a), the FC magnetization data indeed show a thermal hysteresis in the vicinity of $T_{\rm m}$, demonstrating the nature of first-order transition. In the cooling process, Meissner state is first stabilized, which expectedly gives a lower value of magnetization because of Meissner effect~\cite{note2}. On the other hand, In the FCH process from $T_{\rm m}$ to $T_{\rm sc}$, some ``superheated" spontaneous vortices survive accompanying with the ``polarization" of Eu spins until $T_{\rm sc}$, which gives rise to a higher magnetization value. The hysteresis regime extends up to $T_{\rm sc}$, suggesting that the SV state could be stabilized by the Eu-spin ferromagnetic fluctuations. The magnetization differences of the cooling and warming data, $\Delta M_c = M_{\mathrm{FCC}} - M_{\mathrm{FCH}}$, are plotted in Fig.~\ref{fig3}(b). One sees that $\Delta M_c$ drops at $T_{\rm sc}$, and it increases rapidly till $T_{\rm m}$. The maximum of $|\Delta M_c|$ increases with the applied field. Fig.~\ref{fig3}(c) plots the $\Delta M_c$ value at $T_{\rm m}$ ($\Delta M_{c}^{T_{\rm m}}$) as a function of external field. Remarkably, $\Delta M_{c}^{T_{\rm m}}$ is exactly proportional to the field (note that the field accuracy is self-checked by the field-dependent magnetization at 30 K shown on the right axis). In fact, $\Delta M_c$ can be fully scaled with the applied field, as shown in Fig.~\ref{fig3}(d). Here we emphasize that the thermal hysteresis is always observable, even at very low magnetic fields, for different pieces of the sample. By contrast, no thermal hysteresis is seen in overdoped samples where only a ferromagnetic transition takes place. This further rules out the possibility that the domain-wall depinning could be responsible for the large thermal hysteresis.

The magnetization difference at $T_{\rm m}$, $\Delta M_{c}^{T_{\rm m}}$, can be understood as follows. For $T\rightarrow T_{\rm m}^{-}$ (FCH data), the SV state dominates. The magnetic contribution of SVs is always accompanied with the ferromagnetic domains. Owing to the existence of multi-domain, the magnetic fluxes from SVs cancels out at zero field \cite{UCoGe2012}, and with applying fields, the moment appears to be proportional to the external field. When temperature exceeds $T_{\rm m}$, SVs still survive (superheating effect) although the FM vanishes. Namely, the external magnetic field penetrates the sample where superconducting layers contain SVs and, the Eu$^{2+}$ spins are basically in the Curie-Weiss paramagnetic state [see the two right-side cartoons in Fig.~\ref{fig3}(a)]. Thus, the FCH magnetic susceptibility at $T_{\rm m}$ is approximately equal to the Curie-Weiss paramagnetic susceptibility, i.e., $\chi_{\mathrm{FCH}}^{T_{\rm m}} \approx \chi_{\mathrm{CW}}^{T_{\rm m}}$. For $T\rightarrow T_{\rm m}^{+}$ (FCC data), on the other hand, the Meissner state dominates, which gives an additional diamagnetic susceptibility of $\chi_{\mathrm{sc}}^{T_{\rm m}}$, yielding $\chi_{\mathrm{FCC}}^{T_{\rm m}} \approx \chi_{\mathrm{CW}}^{T_{\rm m}} + \chi_{\mathrm{sc}}^{T_{\rm m}}$. Thus we have, $\Delta \chi_c^{T_{\rm m}} = \chi_{\mathrm{FCC}}^{T_{\rm m}} - \chi_{\mathrm{FCH}}^{T_{\rm m}} \approx \chi_{\mathrm{sc}}^{T_{\rm m}}$, which simply reflects the superconducting magnetic expulsion (see the cartoon pictures). The Meissner volume fraction can be estimated to be, $4\pi\Delta M_{c}^{T_{\rm m}}/H \approx 15\%$, which is not surprising because of the unavoidable flux pinning effect.

\begin{figure}
\includegraphics[width=7cm]{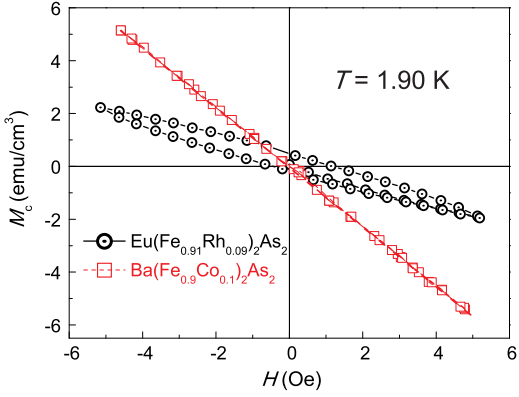}
\caption{\textbf{Isothermal magnetization curves at 1.90 K under magnetic fields parallel to the $c$ axis for Eu(Fe$_{0.91}$Rh$_{0.09})_{2}$As$_{2}$ and Ba(Fe$_{0.9}$Co$_{0.1})_{2}$As$_{2}$.}}
\label{fig4}
\end{figure}

Above we demonstrate the first-order transition from a Meissner state to an SV phase with decreasing temperature. This suggests that the SV phase represents the ground state in Eu(Fe$_{0.91}$Rh$_{0.09}$)$_{2}$As$_{2}$. If this is the case, one expects that the lower critical field at zero temperature, $H_{\mathrm{c1}}(0)$, would be zero~\cite{Varma3,UCoGe-sv,UCoGe2012}. Fig.~\ref{fig4} shows the low-temperature isothermal magnetization, $M_{c}(H)$, for Eu(Fe$_{0.91}$Rh$_{0.09}$)$_{2}$As$_{2}$, in comparison with that of the nonmagnetic superconducting analog, Ba(Fe$_{0.9}$Co$_{0.1}$)$_{2}$As$_{2}$. The latter shows an essentially linear $M_{c}(H)$ since the applied fields are much lower than the $H_{\mathrm{c1}}(0)$. In contrast, Eu(Fe$_{0.91}$Rh$_{0.09}$)$_{2}$As$_{2}$ displays a non-linear virgin $M_{c}(H)$ curve and an obvious magnetic hysteresis loop. This means that, in addition to the superconducting magnetic shielding effect, the external field always penetrates the sample, even if the field is around zero. In other words, Eu(Fe$_{0.91}$Rh$_{0.09}$)$_{2}$As$_{2}$ is intrinsically in a mixed state below $T_{\rm m}$.

\begin{figure}
\includegraphics[width=7cm]{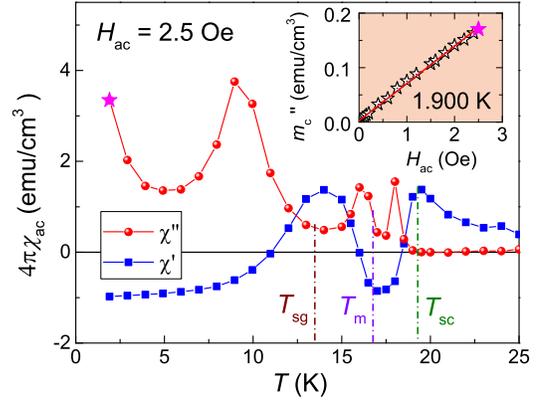}
\caption{\textbf{Temperature dependence of real and imaginary parts ($\chi'$ and $\chi''$) of ac susceptibility at zero dc magnetic field.} The amplitude of the driving ac magnetic field (along the $c$ axis) is, $H_{\mathrm{ac}}$ = 2.5 Oe. The demagnetization effect has been taken into consideration. $T_{\rm sc}$, $T_{\rm m}$, and $T_{\rm sg}$ are the superconducting, ferromagnetic, and spin-glass transition temperatures, respectively. The inset plots the imaginary part of the ac magnetization at 1.9 K as a function of $H_{\mathrm{ac}}$. The solid line is the linear fit.}
\label{fig5}
\end{figure}

Another piece of evidence for the mixed state at zero field comes from the ac magnetic susceptibility measurements. As shown in the main panel of Fig.~\ref{fig5}, one can clearly distinguish $T_{\rm sc}$ and $T_{\rm m}$ from the real part of the ac susceptibility, $\chi'$. The magnetic shielding effect below $T_{\rm sc}$ is much more obvious than that of the dc magnetic measurement above. The imaginary part of the susceptibility, $\chi''$, which is sensitive to dissipations, shows two sharp peaks below $T_{\rm sc}$ and $T_{\rm m}$, respectively. An additional large broad peak appears below the re-entrant spin-glass temperature $T_{\rm sg} \approx$ 13.5 K \cite{Eu122Rh}. Note that this $\chi''$ peak may also be contributed from the SV liquid-to-solid transition.

Remarkably, the $\chi''$ value at the lowest temperature of 1.90 K in our measurements remains considerably high at the driving field $H_{\mathrm{ac}}$ = 2.5 Oe, verifying that it is in a mixed state. To examine if there is a lower limit of the ac field, we performed a field-dependent ac magnetization measurement, the imaginary part ($m_{c}''$) of which is shown in the inset of Fig.~\ref{fig5}. One sees that $m_{c}''$ is exactly proportional to $H_{\mathrm{ac}}$. According to the critical-state model \cite{bean}, $m_{c}''$ will be zero for $H_{\mathrm{ac}} < H_{\rm c1}$; while $m_{c}''$ = $\beta$($H_{\mathrm{ac}} - H_{\rm c1})^{2}/H_{\mathrm{ac}}$ ($\beta$ is the sample's geometrical factor) for $H_{\mathrm{ac}} > H_{\rm c1}$. Both the non-zero $m_{c}''$ and the linearity of $m_{c}''(H_{\mathrm{ac}})$ through the origin indicate that $H_{\rm c1}$ must be zero. Similar observation is seen in the SFM UCoGe \cite{UCoGe2012}.

\

\textbf{Discussion}

The above results allow us to arrive at the following picture for the Eu(Fe$_{0.91}$Rh$_{0.09}$)$_{2}$As$_{2}$ FSC in the absence of external magnetic field. At $T>T_{\rm sc}$, the [(Fe,Rh)$_{2}$As$_{2}$]$^{2-}$ and Eu$^{2+}$ layers are Pauli paramagnetic and Curie-Weiss paramagnetic, respectively. When cooled below $T_{\rm sc}$, the [(Fe,Rh)$_{2}$As$_{2}$]$^{2-}$ layers become superconducting, showing a Meissner state coexisting with the Curie-Weiss paramagnetism of Eu$^{2+}$ spins. With further decreasing temperature to below $T_{\rm m}$, the Eu$^{2+}$ spins are ferromagnetic ordered along the $c$ axis, which generates an internal field far above the expected $H_{\mathrm{c1}}^{*}(0)$. Superconducting vortices then form spontaneously in the [(Fe,Rh)$_{2}$As$_{2}$]$^{2-}$ layers. In the temperature range of 8 K $<T<T_{\rm m}$, the SVs are mobile, which leads to the revival of resistance. The subsequent solidification of the SVs below 8 K gives rise to the zero-resistance state. Therefore, the ground state is an SV solid, which reconciles SC and FM in Eu(Fe$_{0.91}$Rh$_{0.09}$)$_{2}$As$_{2}$.

Finally we note that, apart from the formation of the SV phase, an alternative that allows to reconcile the SC and FM is the so-called Fulde-Ferrell-Larkin-Ovchinnikov (FFLO) state characterized by a spatial modulation of the superconducting order parameter~\cite{FF,LO}. Nevertheless, in general, realization of an FFLO state at zero external field needs more rigorous conditions. Among them are Pauli-limited $H_{\mathrm{c2}}^{\perp}$ (for $H\perp ab$) with a large Maki parameter and clean limit for the SC, which cannot be satisfied in the present system. The $H_{\mathrm{c2}}^{\perp}(T)$ curve in Eu(Fe$_{0.91}$Rh$_{0.09}$)$_{2}$As$_{2}$ keeps linear down to 0.2$T_{\rm sc}$~\cite{Eu122Rh}, indicating that the orbital-limiting effect dominates. Besides, the large residual resistivity ($\sim$60 $\mu \Omega$ cm) as well as the small residual resistivity ratio (RRR = 2.6)~\cite{Eu122Rh} suggests a dirty limit. Both properties actually favor the SV scenario. Nevertheless, here we note that the recently discovered 1144-type FSC~\cite{Rb1144,Cs1144} could be the candidate for an FFLO state, because their nonmagnetic analog, CaKFe$_4$As$_4$, indeed shows a large Maki parameter together with a clean limit for the SC~\cite{canfield}.

In summary, we have studied the low-field magnetic properties for the iron-based ferromagnetic superconductor Eu(Fe$_{0.91}$Rh$_{0.09}$)$_{2}$As$_{2}$. We observed a remarkable thermal hysteresis around the ferromagnetic transition in the superconducting state, even under a vanishingly small field, demonstrating the unique first-order transition from a Meissner state to an SV phase. The SV ground state is further corroborated by the non-linear virgin dc magnetization as well as the non-zero imaginary part of ac magnetic susceptibility under extremely low external fields at $T \ll T_\mathrm{sc}$. The unambiguous demonstration of the SV ground state in the iron-based FSC lays a solid foundation for future studies. For example, it is of great interest to see whether the SV solid behaves like a glassy or a lattice state. The imaging observations such as magnetic-force microscopy as well as the small-angle neutron scattering technique may help to clarify this interesting issue.
\\

\textbf{Methods}
\\
\textbf{Crystal growth.} High-quality crystals of Eu(Fe$_{0.91}$Rh$_{0.09}$)$_{2}$As$_{2}$ were grown by a self-flux method \cite{Eu122Rh,Eu122Ir-jwh}. First, mixtures of Eu (99.9\%), Fe (99.998\%), Rh (99.9\%), and As (99.999\%) powders in a molar ratio of Eu:Fe:Rh:As = 1:4.4:0.6:5 reacted at 973 K for 24 h in a sealed evacuated quartz ampoule. The precursor was ground, and then was loaded into an alumina crucible. The crucible was sealed in a stainless steel tube by arc welding under an atmosphere of argon. The assembly was subsequently heated up to 1573 K and, holding for 5 h, in a muffle furnace with the flow of argon gas. The crystal growth took place during the slow cooling down to 1223 K at the rate of 4 K/h. Large crystals with typical size of 3$\times$3$\times$0.5 mm$^3$ were harvested.

\textbf{Structural and compositional characterizations.} We checked the as-grown crystal flakes by x-ray diffraction using a PANAlytical x-ray diffractometer (using Cu $K_{\alpha1}$ monochromatic radiation) at room temperature. All the crystals show only (00$l$) reflections with even $l$ values, similar to the previous report \cite{Eu122Rh}. The $c$ axis is then determined to be 12.016(1) \AA. The crystal structure is analogous to EuFe$_2$As$_2$ ($c$ = 12.136 \AA) \cite{Eu122-rz}, yet it consists of superconducting [(Fe,Rh)$_{2}$As$_{2}$]$^{2-}$ layers separated by magnetic Eu$^{2+}$ ions. The full width at half maximum (FWHM) of the reflection peaks is typically 2$\theta$ = 0.06$^{\circ}$, verifying the high quality of the crystals. The real composition of the crystal was determined by energy dispersive x-ray spectroscopy, which gives the chemical formula of Eu(Fe$_{0.91}$Rh$_{0.09}$)$_{2}$As$_{2}$.

\textbf{Physical properties.} The electrical and magnetic properties of the Eu(Fe$_{0.91}$Rh$_{0.09}$)$_{2}$As$_{2}$ crystals were reported previously \cite{Eu122Rh}, which demonstrate a superconducting transition at $T_{\mathrm{sc}}$ = 19.6 K, followed by a ferromagnetic transition at $T_{\mathrm{m}}$ = 16.8 K. The isothermal magnetization loops below $T_{\mathrm{m}}$ show characteristic features for both FM and SC. The saturation magnetization achieves $M_{\mathrm{sat}} = 6.5$ $\mu_{\mathrm{B}}/\mathrm{Eu}$, confirming that the Eu spins align ferromagnetically.

\textbf{Low-field magnetic measurements.} We selected a free-standing crystal for all the measurements in this paper. Magnetic measurements were carried out on a Quantum Design Magnetic Property Measurement System. The residual field in the superconducting magnet, after being removed by a degaussing procedure prior to the measurements, is less than $\pm$0.05 Oe. The crystal was carefully mounted into the sample holder with the applied field perpendicular to the crystal plate, such that the external field is either parallel or antiparallel to the internal field. The FC data were collected in both heating and cooling procedures. In the ac susceptibility measurement, the frequency was set to 1.0 Hz. The demagnetization effect is taken into account on the basis of the sample's geometry in respect to the field direction.


\bibliography{SV}

\begin{thebibliography}{40}%
\makeatletter
\providecommand \@ifxundefined [1]{%
 \@ifx{#1\undefined}
}%
\providecommand \@ifnum [1]{%
 \ifnum #1\expandafter \@firstoftwo
 \else \expandafter \@secondoftwo
 \fi
}%
\providecommand \@ifx [1]{%
 \ifx #1\expandafter \@firstoftwo
 \else \expandafter \@secondoftwo
 \fi
}%
\providecommand \natexlab [1]{#1}%
\providecommand \enquote  [1]{``#1''}%
\providecommand \bibnamefont  [1]{#1}%
\providecommand \bibfnamefont [1]{#1}%
\providecommand \citenamefont [1]{#1}%
\providecommand \href@noop [0]{\@secondoftwo}%
\providecommand \href[0]{\begingroup \@sanitize@url \@href}%
\providecommand \@href[1]{\@@startlink{#1}\@@href}%
\providecommand \@@href[1]{\endgroup#1\@@endlink}%
\providecommand \@sanitize@url [0]{\catcode `\\12\catcode `\$12\catcode
  `\&12\catcode `\#12\catcode `\^12\catcode `\_12\catcode `\%12\relax}%
\providecommand \@@startlink[1]{}%
\providecommand \@@endlink[0]{}%
\providecommand \url  [0]{\begingroup\@sanitize@url \@url }%
\providecommand \@url [1]{\endgroup\@href {#1}{\urlprefix }}%
\providecommand \urlprefix  [0]{URL }%
\providecommand \Eprint [0]{\href }%
\providecommand \doibase [0]{http://dx.doi.org/}%
\providecommand \selectlanguage [0]{\@gobble}%
\providecommand \bibinfo  [0]{\@secondoftwo}%
\providecommand \bibfield  [0]{\@secondoftwo}%
\providecommand \translation [1]{[#1]}%
\providecommand \BibitemOpen [0]{}%
\providecommand \bibitemStop [0]{}%
\providecommand \bibitemNoStop [0]{.\EOS\space}%
\providecommand \EOS [0]{\spacefactor3000\relax}%
\providecommand \BibitemShut  [1]{\csname bibitem#1\endcsname}%
\let\auto@bib@innerbib\@empty
\bibitem [{\citenamefont {Blount}\ and\ \citenamefont {Varma}(1979)}]{Varma1}%
  \BibitemOpen
  \bibfield  {author} {\bibinfo {author} {\bibfnamefont {E.~I.}\ \bibnamefont
  {Blount}}\ and\ \bibinfo {author} {\bibfnamefont {C.~M.}\ \bibnamefont
  {Varma}},\ }\enquote {\bibinfo {title} {Electromagnetic Effects near the
  Superconductor-to-Ferromagnet Transition},}\ \href{\doibase
  10.1103/PhysRevLett.42.1079} {\bibfield  {journal} {\bibinfo  {journal}
  {Phys. Rev. Lett.}\ }\textbf {\bibinfo {volume} {42}},\ \bibinfo {pages}
  {1079} (\bibinfo {year} {1979})}\BibitemShut {NoStop}%
\bibitem [{\citenamefont {Tachiki}\ \emph {et~al.}(1980)\citenamefont
  {Tachiki}, \citenamefont {Matsumoto}, \citenamefont {Koyama},\ and\
  \citenamefont {Umezawa}}]{Tachiki}%
  \BibitemOpen
  \bibfield  {author} {\bibinfo {author} {\bibfnamefont {M.}~\bibnamefont
  {Tachiki}}, \bibinfo {author} {\bibfnamefont {H.}~\bibnamefont {Matsumoto}},
  \bibinfo {author} {\bibfnamefont {T.}~\bibnamefont {Koyama}}, \ and\ \bibinfo
  {author} {\bibfnamefont {H.}~\bibnamefont {Umezawa}},\ }\enquote {\bibinfo
  {title} {Self-induced vortices in magnetic superconductors},}\ \href{\doibase
  http://dx.doi.org/10.1016/0038-1098(80)90620-1} {\bibfield  {journal}
  {\bibinfo  {journal} {Solid State Commun.}\ }\textbf {\bibinfo {volume}
  {34}},\ \bibinfo {pages} {19} (\bibinfo {year} {1980})}\BibitemShut {NoStop}%
\bibitem [{\citenamefont {Greenside}\ \emph {et~al.}(1981)\citenamefont
  {Greenside}, \citenamefont {Blount},\ and\ \citenamefont {Varma}}]{Varma2}%
  \BibitemOpen
  \bibfield  {author} {\bibinfo {author} {\bibfnamefont {H.~S.}\ \bibnamefont
  {Greenside}}, \bibinfo {author} {\bibfnamefont {E.~I.}\ \bibnamefont
  {Blount}}, \ and\ \bibinfo {author} {\bibfnamefont {C.~M.}\ \bibnamefont
  {Varma}},\ }\enquote {\bibinfo {title} {Possible Coexisting Superconducting
  and Magnetic States},}\ \href{\doibase 10.1103/PhysRevLett.46.49} {\bibfield
  {journal} {\bibinfo  {journal} {Phys. Rev. Lett.}\ }\textbf {\bibinfo
  {volume} {46}},\ \bibinfo {pages} {49} (\bibinfo {year} {1981})}\BibitemShut
  {NoStop}%
\bibitem [{\citenamefont {Kuper}\ \emph {et~al.}(1980)\citenamefont {Kuper},
  \citenamefont {Revzen},\ and\ \citenamefont {Ron}}]{Kuper}%
  \BibitemOpen
  \bibfield  {author} {\bibinfo {author} {\bibfnamefont {C.~G.}\ \bibnamefont
  {Kuper}}, \bibinfo {author} {\bibfnamefont {M.}~\bibnamefont {Revzen}}, \
  and\ \bibinfo {author} {\bibfnamefont {A.}~\bibnamefont {Ron}},\ }\enquote
  {\bibinfo {title} {Ferromagnetic Superconductors: A Vortex Phase in Ternary
  Rare-Earth Compounds},}\ \href{\doibase 10.1103/PhysRevLett.44.1545}
  {\bibfield  {journal} {\bibinfo  {journal} {Phys. Rev. Lett.}\ }\textbf
  {\bibinfo {volume} {44}},\ \bibinfo {pages} {1545} (\bibinfo {year}
  {1980})}\BibitemShut {NoStop}%
\bibitem [{\citenamefont {Radzihovsky}\ \emph {et~al.}(2001)\citenamefont
  {Radzihovsky}, \citenamefont {Ettouhami}, \citenamefont {Saunders},\ and\
  \citenamefont {Toner}}]{svglass}%
  \BibitemOpen
  \bibfield  {author} {\bibinfo {author} {\bibfnamefont {L.}~\bibnamefont
  {Radzihovsky}}, \bibinfo {author} {\bibfnamefont {A.~M.}\ \bibnamefont
  {Ettouhami}}, \bibinfo {author} {\bibfnamefont {K.}~\bibnamefont {Saunders}},
  \ and\ \bibinfo {author} {\bibfnamefont {J.}~\bibnamefont {Toner}},\
  }\enquote {\bibinfo {title} {``Soft'' Anharmonic Vortex Glass in
  Ferromagnetic Superconductors},}\ \href{\doibase
  10.1103/PhysRevLett.87.027001} {\bibfield  {journal} {\bibinfo  {journal}
  {Phys. Rev. Lett.}\ }\textbf {\bibinfo {volume} {87}},\ \bibinfo {pages}
  {027001} (\bibinfo {year} {2001})}\BibitemShut {NoStop}%
\bibitem [{\citenamefont {Weng}\ and\ \citenamefont {Muthukumar}(2002)}]{wzy}%
  \BibitemOpen
  \bibfield  {author} {\bibinfo {author} {\bibfnamefont {Z.~Y.}\ \bibnamefont
  {Weng}}\ and\ \bibinfo {author} {\bibfnamefont {V.~N.}\ \bibnamefont
  {Muthukumar}},\ }\enquote {\bibinfo {title} {Spontaneous vortex phase in the
  bosonic resonating valence bond theory},}\ \href{\doibase
  10.1103/PhysRevB.66.094509} {\bibfield  {journal} {\bibinfo  {journal} {Phys.
  Rev. B}\ }\textbf {\bibinfo {volume} {66}},\ \bibinfo {pages} {094509}
  (\bibinfo {year} {2002})}\BibitemShut {NoStop}%
\bibitem [{\citenamefont {Bulaevskii}\ \emph {et~al.}(1985)\citenamefont
  {Bulaevskii}, \citenamefont {Buzdin}, \citenamefont {Kuli\'{c}},\ and\
  \citenamefont {Panjukov}}]{Buzdin85}%
  \BibitemOpen
  \bibfield  {author} {\bibinfo {author} {\bibfnamefont {L.}~\bibnamefont
  {Bulaevskii}}, \bibinfo {author} {\bibfnamefont {A.}~\bibnamefont {Buzdin}},
  \bibinfo {author} {\bibfnamefont {M.}~\bibnamefont {Kuli\'{c}}}, \ and\
  \bibinfo {author} {\bibfnamefont {S.}~\bibnamefont {Panjukov}},\ }\enquote
  {\bibinfo {title} {Coexistence of superconductivity and magnetism theoretical
  predictions and experimental results},}\ \href{\doibase
  10.1080/00018738500101741} {\bibfield  {journal} {\bibinfo  {journal} {Adv.
  Phys.}\ }\textbf {\bibinfo {volume} {34}},\ \bibinfo {pages} {175} (\bibinfo
  {year} {1985})}\BibitemShut {NoStop}%
\bibitem [{\citenamefont {Sonin}\ and\ \citenamefont {Felner}(1998)}]{Felner}%
  \BibitemOpen
  \bibfield  {author} {\bibinfo {author} {\bibfnamefont {E.~B.}\ \bibnamefont
  {Sonin}}\ and\ \bibinfo {author} {\bibfnamefont {I.}~\bibnamefont {Felner}},\
  }\enquote {\bibinfo {title} {Spontaneous vortex phase in a superconducting
  weak ferromagnet},}\ \href{\doibase 10.1103/PhysRevB.57.R14000} {\bibfield
  {journal} {\bibinfo  {journal} {Phys. Rev. B}\ }\textbf {\bibinfo {volume}
  {57}},\ \bibinfo {pages} {14000(R)} (\bibinfo {year} {1998})}\BibitemShut
  {NoStop}%
\bibitem [{\citenamefont {Bernhard}\ \emph {et~al.}(2000)\citenamefont
  {Bernhard}, \citenamefont {Tallon}, \citenamefont {Br\"ucher},\ and\
  \citenamefont {Kremer}}]{Bernhard}%
  \BibitemOpen
  \bibfield  {author} {\bibinfo {author} {\bibfnamefont {C.}~\bibnamefont
  {Bernhard}}, \bibinfo {author} {\bibfnamefont {J.~L.}\ \bibnamefont
  {Tallon}}, \bibinfo {author} {\bibfnamefont {E.}~\bibnamefont {Br\"ucher}}, \
  and\ \bibinfo {author} {\bibfnamefont {R.~K.}\ \bibnamefont {Kremer}},\
  }\enquote {\bibinfo {title} {Evidence for a bulk Meissner state in the
  ferromagnetic superconductor
  ${\mathrm{RuSr}}_{2}{\mathrm{GdCu}}_{2}{\mathrm{O}}_{8}$ from dc
  magnetization},}\ \href{\doibase 10.1103/PhysRevB.61.R14960} {\bibfield
  {journal} {\bibinfo  {journal} {Phys. Rev. B}\ }\textbf {\bibinfo {volume}
  {61}},\ \bibinfo {pages} {14960(R)} (\bibinfo {year} {2000})}\BibitemShut
  {NoStop}%
\bibitem [{\citenamefont {Deguchi}\ \emph {et~al.}(2010)\citenamefont
  {Deguchi}, \citenamefont {Osaki}, \citenamefont {Ban}, \citenamefont
  {Tamura}, \citenamefont {Simura}, \citenamefont {Sakakibara}, \citenamefont
  {Satoh},\ and\ \citenamefont {Sato}}]{UCoGe-sv}%
  \BibitemOpen
  \bibfield  {author} {\bibinfo {author} {\bibfnamefont {K.}~\bibnamefont
  {Deguchi}}, \bibinfo {author} {\bibfnamefont {E.}~\bibnamefont {Osaki}},
  \bibinfo {author} {\bibfnamefont {S.}~\bibnamefont {Ban}}, \bibinfo {author}
  {\bibfnamefont {N.}~\bibnamefont {Tamura}}, \bibinfo {author} {\bibfnamefont
  {Y.}~\bibnamefont {Simura}}, \bibinfo {author} {\bibfnamefont
  {T.}~\bibnamefont {Sakakibara}}, \bibinfo {author} {\bibfnamefont
  {I.}~\bibnamefont {Satoh}}, \ and\ \bibinfo {author} {\bibfnamefont {N.~K.}\
  \bibnamefont {Sato}},\ }\enquote {\bibinfo {title} {Absence of Meissner State
  and Robust Ferromagnetism in the Superconducting State of UCoGe: Possible
  Evidence of Spontaneous Vortex State},}\ \href{\doibase
  10.1143/JPSJ.79.083708} {\bibfield  {journal} {\bibinfo  {journal} {J. Phys.
  Soc. Jpn.}\ }\textbf {\bibinfo {volume} {79}},\ \bibinfo {pages} {083708}
  (\bibinfo {year} {2010})}\BibitemShut {NoStop}%
\bibitem [{\citenamefont {Paulsen}\ \emph {et~al.}(2012)\citenamefont
  {Paulsen}, \citenamefont {Hykel}, \citenamefont {Hasselbach},\ and\
  \citenamefont {Aoki}}]{UCoGe2012}%
  \BibitemOpen
  \bibfield  {author} {\bibinfo {author} {\bibfnamefont {C.}~\bibnamefont
  {Paulsen}}, \bibinfo {author} {\bibfnamefont {D.~J.}\ \bibnamefont {Hykel}},
  \bibinfo {author} {\bibfnamefont {K.}~\bibnamefont {Hasselbach}}, \ and\
  \bibinfo {author} {\bibfnamefont {D.}~\bibnamefont {Aoki}},\ }\enquote
  {\bibinfo {title} {Observation of the Meissner-Ochsenfeld Effect and the
  Absence of the Meissner State in UCoGe},}\ \href{\doibase
  10.1103/PhysRevLett.109.237001} {\bibfield  {journal} {\bibinfo  {journal}
  {Phys. Rev. Lett.}\ }\textbf {\bibinfo {volume} {109}},\ \bibinfo {pages}
  {237001} (\bibinfo {year} {2012})}\BibitemShut {NoStop}%
\bibitem [{\citenamefont {Ng}\ and\ \citenamefont {Varma}(1997)}]{Varma3}%
  \BibitemOpen
  \bibfield  {author} {\bibinfo {author} {\bibfnamefont {T.~K.}\ \bibnamefont
  {Ng}}\ and\ \bibinfo {author} {\bibfnamefont {C.~M.}\ \bibnamefont {Varma}},\
  }\enquote {\bibinfo {title} {Spontaneous Vortex Phase Discovered?}}\
  \href{\doibase 10.1103/PhysRevLett.78.330} {\bibfield  {journal} {\bibinfo
  {journal} {Phys. Rev. Lett.}\ }\textbf {\bibinfo {volume} {78}},\ \bibinfo
  {pages} {330} (\bibinfo {year} {1997})}\BibitemShut {NoStop}%
\bibitem [{\citenamefont {Chia}\ \emph {et~al.}(2006)\citenamefont {Chia},
  \citenamefont {Salamon}, \citenamefont {Park}, \citenamefont {Kim},
  \citenamefont {Lee},\ and\ \citenamefont {Takeya}}]{ErNi2B2C-sv}%
  \BibitemOpen
  \bibfield  {author} {\bibinfo {author} {\bibfnamefont {E.~E.~M.}\
  \bibnamefont {Chia}}, \bibinfo {author} {\bibfnamefont {M.~B.}\ \bibnamefont
  {Salamon}}, \bibinfo {author} {\bibfnamefont {T.}~\bibnamefont {Park}},
  \bibinfo {author} {\bibfnamefont {H.-J.}\ \bibnamefont {Kim}}, \bibinfo
  {author} {\bibfnamefont {S.-I.}\ \bibnamefont {Lee}}, \ and\ \bibinfo
  {author} {\bibfnamefont {H.}~\bibnamefont {Takeya}},\ }\enquote {\bibinfo
  {title} {Observation of the spontaneous vortex phase in the weakly
  ferromagnetic superconductor {ErNi$_2$B$_2$C}: A penetration depth study},}\
  \href{http://stacks.iop.org/0295-5075/73/i=5/a=772} {\bibfield  {journal}
  {\bibinfo  {journal} {EPL (Europhysics Letters)}\ }\textbf {\bibinfo {volume}
  {73}},\ \bibinfo {pages} {772} (\bibinfo {year} {2006})}\BibitemShut
  {NoStop}%
\bibitem [{\citenamefont {Papageorgiou}\ \emph {et~al.}(2006)\citenamefont
  {Papageorgiou}, \citenamefont {Casini}, \citenamefont {Braun}, \citenamefont
  {Herrmannsd{\"o}rfer}, \citenamefont {Bianchi},\ and\ \citenamefont
  {Wosnitza}}]{Papageorgiou}%
  \BibitemOpen
  \bibfield  {author} {\bibinfo {author} {\bibfnamefont {T.~P.}\ \bibnamefont
  {Papageorgiou}}, \bibinfo {author} {\bibfnamefont {E.}~\bibnamefont
  {Casini}}, \bibinfo {author} {\bibfnamefont {H.~F.}\ \bibnamefont {Braun}},
  \bibinfo {author} {\bibfnamefont {T.}~\bibnamefont {Herrmannsd{\"o}rfer}},
  \bibinfo {author} {\bibfnamefont {A.~D.}\ \bibnamefont {Bianchi}}, \ and\
  \bibinfo {author} {\bibfnamefont {J.}~\bibnamefont {Wosnitza}},\ }\enquote
  {\bibinfo {title} {Magnetization, vortex state and specific heat in the
  superconducting state of
  ${\mathrm{RuSr}}_{2}{\mathrm{GdCu}}_{2}{\mathrm{O}}_{8}$},}\ \href{\doibase
  10.1140/epjb/e2006-00317-4} {\bibfield  {journal} {\bibinfo  {journal} {The
  European Physical Journal B - Condensed Matter and Complex Systems}\ }\textbf
  {\bibinfo {volume} {52}},\ \bibinfo {pages} {383} (\bibinfo {year}
  {2006})}\BibitemShut {NoStop}%
\bibitem [{\citenamefont {Zapf}\ and\ \citenamefont {Dressel}(2017)}]{Dressel}%
  \BibitemOpen
  \bibfield  {author} {\bibinfo {author} {\bibfnamefont {S.}~\bibnamefont
  {Zapf}}\ and\ \bibinfo {author} {\bibfnamefont {M.}~\bibnamefont {Dressel}},\
  }\enquote {\bibinfo {title} {Europium-based iron pnictides: a unique
  laboratory for magnetism, superconductivity and structural effects},}\
  \href{http://stacks.iop.org/0034-4885/80/i=1/a=016501} {\bibfield  {journal}
  {\bibinfo  {journal} {Reports on Progress in Physics}\ }\textbf {\bibinfo
  {volume} {80}},\ \bibinfo {pages} {016501} (\bibinfo {year}
  {2017})}\BibitemShut {NoStop}%
\bibitem [{\citenamefont {Ren}\ \emph {et~al.}(2009)\citenamefont {Ren},
  \citenamefont {Tao}, \citenamefont {Jiang}, \citenamefont {Feng},
  \citenamefont {Wang}, \citenamefont {Dai}, \citenamefont {Cao},\ and\
  \citenamefont {Xu}}]{Eu122P-rz}%
  \BibitemOpen
  \bibfield  {author} {\bibinfo {author} {\bibfnamefont {Z.}~\bibnamefont
  {Ren}}, \bibinfo {author} {\bibfnamefont {Q.}~\bibnamefont {Tao}}, \bibinfo
  {author} {\bibfnamefont {S.}~\bibnamefont {Jiang}}, \bibinfo {author}
  {\bibfnamefont {C.}~\bibnamefont {Feng}}, \bibinfo {author} {\bibfnamefont
  {C.}~\bibnamefont {Wang}}, \bibinfo {author} {\bibfnamefont {J.}~\bibnamefont
  {Dai}}, \bibinfo {author} {\bibfnamefont {G.}~\bibnamefont {Cao}}, \ and\
  \bibinfo {author} {\bibfnamefont {Z.}~\bibnamefont {Xu}},\ }\enquote
  {\bibinfo {title} {Superconductivity Induced by Phosphorus Doping and Its
  Coexistence with Ferromagnetism in
  ${\mathrm{EuFe}}_{2}({\mathrm{As}}_{0.7}{\mathrm{P}}_{0.3}{)}_{2}$},}\
  \href{\doibase 10.1103/PhysRevLett.102.137002} {\bibfield  {journal}
  {\bibinfo  {journal} {Phys. Rev. Lett.}\ }\textbf {\bibinfo {volume} {102}},\
  \bibinfo {pages} {137002} (\bibinfo {year} {2009})}\BibitemShut {NoStop}%
\bibitem [{\citenamefont {Jiang}\ \emph {et~al.}(2009)\citenamefont {Jiang},
  \citenamefont {Xing}, \citenamefont {Xuan}, \citenamefont {Ren},
  \citenamefont {Wang}, \citenamefont {Xu},\ and\ \citenamefont
  {Cao}}]{Eu122Co-js}%
  \BibitemOpen
  \bibfield  {author} {\bibinfo {author} {\bibfnamefont {S.}~\bibnamefont
  {Jiang}}, \bibinfo {author} {\bibfnamefont {H.}~\bibnamefont {Xing}},
  \bibinfo {author} {\bibfnamefont {G.}~\bibnamefont {Xuan}}, \bibinfo {author}
  {\bibfnamefont {Z.}~\bibnamefont {Ren}}, \bibinfo {author} {\bibfnamefont
  {C.}~\bibnamefont {Wang}}, \bibinfo {author} {\bibfnamefont {Z.-A.}\
  \bibnamefont {Xu}}, \ and\ \bibinfo {author} {\bibfnamefont {G.}~\bibnamefont
  {Cao}},\ }\enquote {\bibinfo {title} {Superconductivity and local-moment
  magnetism in
  $\text{Eu}{({\text{Fe}}_{0.89}{\text{Co}}_{0.11})}_{2}{\text{As}}_{2}$},}\
  \href{\doibase 10.1103/PhysRevB.80.184514} {\bibfield  {journal} {\bibinfo
  {journal} {Phys. Rev. B}\ }\textbf {\bibinfo {volume} {80}},\ \bibinfo
  {pages} {184514} (\bibinfo {year} {2009})}\BibitemShut {NoStop}%
\bibitem [{\citenamefont {Jiao}\ \emph {et~al.}(2011)\citenamefont {Jiao},
  \citenamefont {Tao}, \citenamefont {Bao}, \citenamefont {Sun}, \citenamefont
  {Feng}, \citenamefont {Xu}, \citenamefont {Nowik}, \citenamefont {Felner},\
  and\ \citenamefont {Cao}}]{Eu122Ru}%
  \BibitemOpen
  \bibfield  {author} {\bibinfo {author} {\bibfnamefont {W.-H.}\ \bibnamefont
  {Jiao}}, \bibinfo {author} {\bibfnamefont {Q.}~\bibnamefont {Tao}}, \bibinfo
  {author} {\bibfnamefont {J.-K.}\ \bibnamefont {Bao}}, \bibinfo {author}
  {\bibfnamefont {Y.-L.}\ \bibnamefont {Sun}}, \bibinfo {author} {\bibfnamefont
  {C.-M.}\ \bibnamefont {Feng}}, \bibinfo {author} {\bibfnamefont {Z.-A.}\
  \bibnamefont {Xu}}, \bibinfo {author} {\bibfnamefont {I.}~\bibnamefont
  {Nowik}}, \bibinfo {author} {\bibfnamefont {I.}~\bibnamefont {Felner}}, \
  and\ \bibinfo {author} {\bibfnamefont {G.-H.}\ \bibnamefont {Cao}},\
  }\enquote {\bibinfo {title} {Anisotropic superconductivity in
  {Eu(Fe$_{0.75}$Ru$_{0.25}$)$_2$As$_2$} ferromagnetic superconductor},}\
  \href{http://stacks.iop.org/0295-5075/95/i=6/a=67007} {\bibfield  {journal}
  {\bibinfo  {journal} {EPL (Europhysics Letters)}\ }\textbf {\bibinfo {volume}
  {95}},\ \bibinfo {pages} {67007} (\bibinfo {year} {2011})}\BibitemShut
  {NoStop}%
\bibitem [{\citenamefont {Jiao}\ \emph {et~al.}(2017)\citenamefont {Jiao},
  \citenamefont {Liu}, \citenamefont {Tang}, \citenamefont {Li}, \citenamefont
  {Xu}, \citenamefont {Ren}, \citenamefont {Xu},\ and\ \citenamefont
  {Cao}}]{Eu122Rh}%
  \BibitemOpen
  \bibfield  {author} {\bibinfo {author} {\bibfnamefont {W.-H.}\ \bibnamefont
  {Jiao}}, \bibinfo {author} {\bibfnamefont {Y.}~\bibnamefont {Liu}}, \bibinfo
  {author} {\bibfnamefont {Z.-T.}\ \bibnamefont {Tang}}, \bibinfo {author}
  {\bibfnamefont {Y.-K.}\ \bibnamefont {Li}}, \bibinfo {author} {\bibfnamefont
  {X.-F.}\ \bibnamefont {Xu}}, \bibinfo {author} {\bibfnamefont
  {Z.}~\bibnamefont {Ren}}, \bibinfo {author} {\bibfnamefont {Z.-A.}\
  \bibnamefont {Xu}}, \ and\ \bibinfo {author} {\bibfnamefont {G.-H.}\
  \bibnamefont {Cao}},\ }\enquote {\bibinfo {title} {Peculiar properties of the
  ferromagnetic superconductor {Eu(Fe$_{0.91}$Rh$_{0.09}$)$_2$As$_2$}},}\
  \href{http://stacks.iop.org/0953-2048/30/i=2/a=025012} {\bibfield  {journal}
  {\bibinfo  {journal} {Supercond. Sci. Technol.}\ }\textbf {\bibinfo {volume}
  {30}},\ \bibinfo {pages} {025012} (\bibinfo {year} {2017})}\BibitemShut
  {NoStop}%
\bibitem [{\citenamefont {Jiao}\ \emph {et~al.}(2013)\citenamefont {Jiao},
  \citenamefont {Zhai}, \citenamefont {Bao}, \citenamefont {Luo}, \citenamefont
  {Tao}, \citenamefont {Feng}, \citenamefont {Xu},\ and\ \citenamefont
  {Cao}}]{Eu122Ir-jwh}%
  \BibitemOpen
  \bibfield  {author} {\bibinfo {author} {\bibfnamefont {W.-H.}\ \bibnamefont
  {Jiao}}, \bibinfo {author} {\bibfnamefont {H.-F.}\ \bibnamefont {Zhai}},
  \bibinfo {author} {\bibfnamefont {J.-K.}\ \bibnamefont {Bao}}, \bibinfo
  {author} {\bibfnamefont {Y.-K.}\ \bibnamefont {Luo}}, \bibinfo {author}
  {\bibfnamefont {Q.}~\bibnamefont {Tao}}, \bibinfo {author} {\bibfnamefont
  {C.-M.}\ \bibnamefont {Feng}}, \bibinfo {author} {\bibfnamefont {Z.-A.}\
  \bibnamefont {Xu}}, \ and\ \bibinfo {author} {\bibfnamefont {G.-H.}\
  \bibnamefont {Cao}},\ }\enquote {\bibinfo {title} {Anomalous critical fields
  and the absence of Meissner state in {Eu(Fe$_{0.88}$Ir$_{0.12}$)$_2$As$_2$}
  crystals},}\ \href{http://stacks.iop.org/1367-2630/15/i=11/a=113002}
  {\bibfield  {journal} {\bibinfo  {journal} {New J. Phys.}\ }\textbf {\bibinfo
  {volume} {15}},\ \bibinfo {pages} {113002} (\bibinfo {year}
  {2013})}\BibitemShut {NoStop}%
\bibitem [{\citenamefont {Paramanik}\ \emph {et~al.}(2013)\citenamefont
  {Paramanik}, \citenamefont {Das}, \citenamefont {Prasad},\ and\ \citenamefont
  {Hossain}}]{Eu122Ir-hossian}%
  \BibitemOpen
  \bibfield  {author} {\bibinfo {author} {\bibfnamefont {U.~B.}\ \bibnamefont
  {Paramanik}}, \bibinfo {author} {\bibfnamefont {D.}~\bibnamefont {Das}},
  \bibinfo {author} {\bibfnamefont {R.}~\bibnamefont {Prasad}}, \ and\ \bibinfo
  {author} {\bibfnamefont {Z.}~\bibnamefont {Hossain}},\ }\enquote {\bibinfo
  {title} {Reentrant superconductivity in
  $\mathrm{Eu}(\mathrm{Fe}_{1-x}\mathrm{Ir}_{x})_{2}\mathrm{As}_{2}$},}\
  \href{http://stacks.iop.org/0953-8984/25/i=26/a=265701} {\bibfield  {journal}
  {\bibinfo  {journal} {J. Phys.: Condens. Matt.}\ }\textbf {\bibinfo {volume}
  {25}},\ \bibinfo {pages} {265701} (\bibinfo {year} {2013})}\BibitemShut
  {NoStop}%
\bibitem [{\citenamefont {Nowik}\ \emph {et~al.}(2011)\citenamefont {Nowik},
  \citenamefont {Felner}, \citenamefont {Ren}, \citenamefont {Cao},\ and\
  \citenamefont {Xu}}]{Felner2011}%
  \BibitemOpen
  \bibfield  {author} {\bibinfo {author} {\bibfnamefont {I.}~\bibnamefont
  {Nowik}}, \bibinfo {author} {\bibfnamefont {I.}~\bibnamefont {Felner}},
  \bibinfo {author} {\bibfnamefont {Z.}~\bibnamefont {Ren}}, \bibinfo {author}
  {\bibfnamefont {G.~H.}\ \bibnamefont {Cao}}, \ and\ \bibinfo {author}
  {\bibfnamefont {Z.~A.}\ \bibnamefont {Xu}},\ }\enquote {\bibinfo {title}
  {Coexistence of ferromagnetism and superconductivity: magnetization and
  Mossbauer studies of EuFe$_{2}$(As$_{1-x}$P$_{x}$)$_{2}$},}\
  \href{http://stacks.iop.org/0953-8984/23/i=6/a=065701} {\bibfield  {journal}
  {\bibinfo  {journal} {J. Phys.: Condens. Matt.}\ }\textbf {\bibinfo {volume}
  {23}},\ \bibinfo {pages} {065701} (\bibinfo {year} {2011})}\BibitemShut
  {NoStop}%
\bibitem [{\citenamefont {Terashima}\ \emph {et~al.}(2009)\citenamefont
  {Terashima}, \citenamefont {Kimata}, \citenamefont {Satsukawa}, \citenamefont
  {Harada}, \citenamefont {Hazama}, \citenamefont {Uji}, \citenamefont
  {Suzuki}, \citenamefont {Matsumoto},\ and\ \citenamefont
  {Murata}}]{terashima}%
  \BibitemOpen
  \bibfield  {author} {\bibinfo {author} {\bibfnamefont {T.}~\bibnamefont
  {Terashima}}, \bibinfo {author} {\bibfnamefont {M.}~\bibnamefont {Kimata}},
  \bibinfo {author} {\bibfnamefont {H.}~\bibnamefont {Satsukawa}}, \bibinfo
  {author} {\bibfnamefont {A.}~\bibnamefont {Harada}}, \bibinfo {author}
  {\bibfnamefont {K.}~\bibnamefont {Hazama}}, \bibinfo {author} {\bibfnamefont
  {S.}~\bibnamefont {Uji}}, \bibinfo {author} {\bibfnamefont {H.~S.}\
  \bibnamefont {Suzuki}}, \bibinfo {author} {\bibfnamefont {T.}~\bibnamefont
  {Matsumoto}}, \ and\ \bibinfo {author} {\bibfnamefont {K.}~\bibnamefont
  {Murata}},\ }\enquote {\bibinfo {title} {EuFe$_2$As$_2$ under High Pressure:
  An Antiferromagnetic Bulk Superconductor},}\ \href{\doibase
  10.1143/JPSJ.78.083701} {\bibfield  {journal} {\bibinfo  {journal} {J. Phys.
  Soc. Jpn.}\ }\textbf {\bibinfo {volume} {78}},\ \bibinfo {pages} {083701}
  (\bibinfo {year} {2009})}\BibitemShut {NoStop}%
\bibitem [{\citenamefont {Jeevan}\ \emph {et~al.}(2011)\citenamefont {Jeevan},
  \citenamefont {Kasinathan}, \citenamefont {Rosner},\ and\ \citenamefont
  {Gegenwart}}]{jeevan}%
  \BibitemOpen
  \bibfield  {author} {\bibinfo {author} {\bibfnamefont {H.~S.}\ \bibnamefont
  {Jeevan}}, \bibinfo {author} {\bibfnamefont {D.}~\bibnamefont {Kasinathan}},
  \bibinfo {author} {\bibfnamefont {H.}~\bibnamefont {Rosner}}, \ and\ \bibinfo
  {author} {\bibfnamefont {P.}~\bibnamefont {Gegenwart}},\ }\enquote {\bibinfo
  {title} {Interplay of antiferromagnetism, ferromagnetism, and
  superconductivity in EuFe$_{2}$(As$_{1-x}$P$_{x}$)$_{2}$ single crystals},}\
  \href{\doibase 10.1103/PhysRevB.83.054511} {\bibfield  {journal} {\bibinfo
  {journal} {Phys. Rev. B}\ }\textbf {\bibinfo {volume} {83}},\ \bibinfo
  {pages} {054511} (\bibinfo {year} {2011})}\BibitemShut {NoStop}%
\bibitem [{\citenamefont {Tokiwa}\ \emph {et~al.}(2012)\citenamefont {Tokiwa},
  \citenamefont {H\"ubner}, \citenamefont {Beck}, \citenamefont {Jeevan},\ and\
  \citenamefont {Gegenwart}}]{tokiwa}%
  \BibitemOpen
  \bibfield  {author} {\bibinfo {author} {\bibfnamefont {Y.}~\bibnamefont
  {Tokiwa}}, \bibinfo {author} {\bibfnamefont {S.-H.}\ \bibnamefont
  {H\"ubner}}, \bibinfo {author} {\bibfnamefont {O.}~\bibnamefont {Beck}},
  \bibinfo {author} {\bibfnamefont {H.~S.}\ \bibnamefont {Jeevan}}, \ and\
  \bibinfo {author} {\bibfnamefont {P.}~\bibnamefont {Gegenwart}},\ }\enquote
  {\bibinfo {title} {Unique phase diagram with narrow superconducting dome in
  EuFe${}_{2}$(As${}_{1\ensuremath{-}x}$P${}_{x}$)${}_{2}$ due to Eu${}^{2+}$
  local magnetic moments},}\ \href{\doibase 10.1103/PhysRevB.86.220505}
  {\bibfield  {journal} {\bibinfo  {journal} {Phys. Rev. B}\ }\textbf {\bibinfo
  {volume} {86}},\ \bibinfo {pages} {220505} (\bibinfo {year}
  {2012})}\BibitemShut {NoStop}%
\bibitem [{\citenamefont {Zapf}\ \emph {et~al.}(2013)\citenamefont {Zapf},
  \citenamefont {Jeevan}, \citenamefont {Ivek}, \citenamefont {Pfister},
  \citenamefont {Klingert}, \citenamefont {Jiang}, \citenamefont {Wu},
  \citenamefont {Gegenwart}, \citenamefont {Kremer},\ and\ \citenamefont
  {Dressel}}]{zapf}%
  \BibitemOpen
  \bibfield  {author} {\bibinfo {author} {\bibfnamefont {S.}~\bibnamefont
  {Zapf}}, \bibinfo {author} {\bibfnamefont {H.~S.}\ \bibnamefont {Jeevan}},
  \bibinfo {author} {\bibfnamefont {T.}~\bibnamefont {Ivek}}, \bibinfo {author}
  {\bibfnamefont {F.}~\bibnamefont {Pfister}}, \bibinfo {author} {\bibfnamefont
  {F.}~\bibnamefont {Klingert}}, \bibinfo {author} {\bibfnamefont
  {S.}~\bibnamefont {Jiang}}, \bibinfo {author} {\bibfnamefont
  {D.}~\bibnamefont {Wu}}, \bibinfo {author} {\bibfnamefont {P.}~\bibnamefont
  {Gegenwart}}, \bibinfo {author} {\bibfnamefont {R.~K.}\ \bibnamefont
  {Kremer}}, \ and\ \bibinfo {author} {\bibfnamefont {M.}~\bibnamefont
  {Dressel}},\ }\enquote {\bibinfo {title}
  {${\mathrm{EuFe}}_{2}({\mathrm{As}}_{1-x}{\mathrm{P}}_{x}{)}_{2}$: Reentrant
  Spin Glass and Superconductivity},}\ \href{\doibase
  10.1103/PhysRevLett.110.237002} {\bibfield  {journal} {\bibinfo  {journal}
  {Phys. Rev. Lett.}\ }\textbf {\bibinfo {volume} {110}},\ \bibinfo {pages}
  {237002} (\bibinfo {year} {2013})}\BibitemShut {NoStop}%
\bibitem [{\citenamefont {Jin}\ \emph {et~al.}(2013)\citenamefont {Jin},
  \citenamefont {Nandi}, \citenamefont {Xiao}, \citenamefont {Su},
  \citenamefont {Zaharko}, \citenamefont {Guguchia}, \citenamefont {Bukowski},
  \citenamefont {Price}, \citenamefont {Jiao}, \citenamefont {Cao},\ and\
  \citenamefont {Br\"uckel}}]{Eu122Co-jwt}%
  \BibitemOpen
  \bibfield  {author} {\bibinfo {author} {\bibfnamefont {W.~T.}\ \bibnamefont
  {Jin}}, \bibinfo {author} {\bibfnamefont {S.}~\bibnamefont {Nandi}}, \bibinfo
  {author} {\bibfnamefont {Y.}~\bibnamefont {Xiao}}, \bibinfo {author}
  {\bibfnamefont {Y.}~\bibnamefont {Su}}, \bibinfo {author} {\bibfnamefont
  {O.}~\bibnamefont {Zaharko}}, \bibinfo {author} {\bibfnamefont
  {Z.}~\bibnamefont {Guguchia}}, \bibinfo {author} {\bibfnamefont
  {Z.}~\bibnamefont {Bukowski}}, \bibinfo {author} {\bibfnamefont
  {S.}~\bibnamefont {Price}}, \bibinfo {author} {\bibfnamefont {W.~H.}\
  \bibnamefont {Jiao}}, \bibinfo {author} {\bibfnamefont {G.~H.}\ \bibnamefont
  {Cao}}, \ and\ \bibinfo {author} {\bibfnamefont {T.}~\bibnamefont
  {Br\"uckel}},\ }\enquote {\bibinfo {title} {Magnetic structure of
  superconducting Eu(Fe${}_{0.82}$Co${}_{0.18}$)${}_{2}$As${}_{2}$ as revealed
  by single-crystal neutron diffraction},}\ \href{\doibase
  10.1103/PhysRevB.88.214516} {\bibfield  {journal} {\bibinfo  {journal} {Phys.
  Rev. B}\ }\textbf {\bibinfo {volume} {88}},\ \bibinfo {pages} {214516}
  (\bibinfo {year} {2013})}\BibitemShut {NoStop}%
\bibitem [{\citenamefont {Nandi}\ \emph
  {et~al.}(2014{\natexlab{a}})\citenamefont {Nandi}, \citenamefont {Jin},
  \citenamefont {Xiao}, \citenamefont {Su}, \citenamefont {Price},
  \citenamefont {Shukla}, \citenamefont {Strempfer}, \citenamefont {Jeevan},
  \citenamefont {Gegenwart},\ and\ \citenamefont {Br\"uckel}}]{Eu122P-xms}%
  \BibitemOpen
  \bibfield  {author} {\bibinfo {author} {\bibfnamefont {S.}~\bibnamefont
  {Nandi}}, \bibinfo {author} {\bibfnamefont {W.~T.}\ \bibnamefont {Jin}},
  \bibinfo {author} {\bibfnamefont {Y.}~\bibnamefont {Xiao}}, \bibinfo {author}
  {\bibfnamefont {Y.}~\bibnamefont {Su}}, \bibinfo {author} {\bibfnamefont
  {S.}~\bibnamefont {Price}}, \bibinfo {author} {\bibfnamefont {D.~K.}\
  \bibnamefont {Shukla}}, \bibinfo {author} {\bibfnamefont {J.}~\bibnamefont
  {Strempfer}}, \bibinfo {author} {\bibfnamefont {H.~S.}\ \bibnamefont
  {Jeevan}}, \bibinfo {author} {\bibfnamefont {P.}~\bibnamefont {Gegenwart}}, \
  and\ \bibinfo {author} {\bibfnamefont {T.}~\bibnamefont {Br\"uckel}},\
  }\enquote {\bibinfo {title} {Coexistence of superconductivity and
  ferromagnetism in P-doped ${\text{EuFe}}_{2}{\mathrm{As}}_{2}$},}\
  \href{\doibase 10.1103/PhysRevB.89.014512} {\bibfield  {journal} {\bibinfo
  {journal} {Phys. Rev. B}\ }\textbf {\bibinfo {volume} {89}},\ \bibinfo
  {pages} {014512} (\bibinfo {year} {2014}{\natexlab{a}})}\BibitemShut
  {NoStop}%
\bibitem [{\citenamefont {Nandi}\ \emph
  {et~al.}(2014{\natexlab{b}})\citenamefont {Nandi}, \citenamefont {Jin},
  \citenamefont {Xiao}, \citenamefont {Su}, \citenamefont {Price},
  \citenamefont {Schmidt}, \citenamefont {Schmalzl}, \citenamefont {Chatterji},
  \citenamefont {Jeevan}, \citenamefont {Gegenwart},\ and\ \citenamefont
  {Br\"uckel}}]{Eu122P-neutron}%
  \BibitemOpen
  \bibfield  {author} {\bibinfo {author} {\bibfnamefont {S.}~\bibnamefont
  {Nandi}}, \bibinfo {author} {\bibfnamefont {W.~T.}\ \bibnamefont {Jin}},
  \bibinfo {author} {\bibfnamefont {Y.}~\bibnamefont {Xiao}}, \bibinfo {author}
  {\bibfnamefont {Y.}~\bibnamefont {Su}}, \bibinfo {author} {\bibfnamefont
  {S.}~\bibnamefont {Price}}, \bibinfo {author} {\bibfnamefont
  {W.}~\bibnamefont {Schmidt}}, \bibinfo {author} {\bibfnamefont
  {K.}~\bibnamefont {Schmalzl}}, \bibinfo {author} {\bibfnamefont
  {T.}~\bibnamefont {Chatterji}}, \bibinfo {author} {\bibfnamefont {H.~S.}\
  \bibnamefont {Jeevan}}, \bibinfo {author} {\bibfnamefont {P.}~\bibnamefont
  {Gegenwart}}, \ and\ \bibinfo {author} {\bibfnamefont {T.}~\bibnamefont
  {Br\"uckel}},\ }\enquote {\bibinfo {title} {Magnetic structure of the
  ${\mathrm{Eu}}^{2+}$ moments in superconducting
  ${\mathrm{EuFe}}_{2}{({\mathrm{As}}_{1\ensuremath{-}x}{\mathrm{P}}_{x})}_{2}$
  with $x=0.19$},}\ \href{\doibase 10.1103/PhysRevB.90.094407} {\bibfield
  {journal} {\bibinfo  {journal} {Phys. Rev. B}\ }\textbf {\bibinfo {volume}
  {90}},\ \bibinfo {pages} {094407} (\bibinfo {year}
  {2014}{\natexlab{b}})}\BibitemShut {NoStop}%
\bibitem [{\citenamefont {Jin}\ \emph {et~al.}(2015)\citenamefont {Jin},
  \citenamefont {Li}, \citenamefont {Su}, \citenamefont {Nandi}, \citenamefont
  {Xiao}, \citenamefont {Jiao}, \citenamefont {Meven}, \citenamefont {Sazonov},
  \citenamefont {Feng}, \citenamefont {Chen}, \citenamefont {Ting},
  \citenamefont {Cao},\ and\ \citenamefont {Br\"uckel}}]{Eu122Ir-jwt}%
  \BibitemOpen
  \bibfield  {author} {\bibinfo {author} {\bibfnamefont {W.~T.}\ \bibnamefont
  {Jin}}, \bibinfo {author} {\bibfnamefont {W.}~\bibnamefont {Li}}, \bibinfo
  {author} {\bibfnamefont {Y.}~\bibnamefont {Su}}, \bibinfo {author}
  {\bibfnamefont {S.}~\bibnamefont {Nandi}}, \bibinfo {author} {\bibfnamefont
  {Y.}~\bibnamefont {Xiao}}, \bibinfo {author} {\bibfnamefont {W.~H.}\
  \bibnamefont {Jiao}}, \bibinfo {author} {\bibfnamefont {M.}~\bibnamefont
  {Meven}}, \bibinfo {author} {\bibfnamefont {A.~P.}\ \bibnamefont {Sazonov}},
  \bibinfo {author} {\bibfnamefont {E.}~\bibnamefont {Feng}}, \bibinfo {author}
  {\bibfnamefont {Y.}~\bibnamefont {Chen}}, \bibinfo {author} {\bibfnamefont
  {C.~S.}\ \bibnamefont {Ting}}, \bibinfo {author} {\bibfnamefont {G.~H.}\
  \bibnamefont {Cao}}, \ and\ \bibinfo {author} {\bibfnamefont
  {T.}~\bibnamefont {Br\"uckel}},\ }\enquote {\bibinfo {title} {Magnetic ground
  state of superconducting
  $\mathrm{Eu}(\mathrm{Fe}{}_{0.88}\mathrm{Ir}{}_{0.12}){}_{2}\mathrm{As}{}_{2}$:
  A combined neutron diffraction and first-principles calculation study},}\
  \href{\doibase 10.1103/PhysRevB.91.064506} {\bibfield  {journal} {\bibinfo
  {journal} {Phys. Rev. B}\ }\textbf {\bibinfo {volume} {91}},\ \bibinfo
  {pages} {064506} (\bibinfo {year} {2015})}\BibitemShut {NoStop}%
\bibitem [{\citenamefont {Abdel-Hafiez}\ \emph {et~al.}(2015)\citenamefont
  {Abdel-Hafiez}, \citenamefont {Zhang}, \citenamefont {He}, \citenamefont
  {Zhao}, \citenamefont {Bergmann}, \citenamefont {Krellner}, \citenamefont
  {Duan}, \citenamefont {Lu}, \citenamefont {Luo}, \citenamefont {Dai},\ and\
  \citenamefont {Chen}}]{Hc1}%
  \BibitemOpen
  \bibfield  {author} {\bibinfo {author} {\bibfnamefont {M.}~\bibnamefont
  {Abdel-Hafiez}}, \bibinfo {author} {\bibfnamefont {Y.}~\bibnamefont {Zhang}},
  \bibinfo {author} {\bibfnamefont {Z.}~\bibnamefont {He}}, \bibinfo {author}
  {\bibfnamefont {J.}~\bibnamefont {Zhao}}, \bibinfo {author} {\bibfnamefont
  {C.}~\bibnamefont {Bergmann}}, \bibinfo {author} {\bibfnamefont
  {C.}~\bibnamefont {Krellner}}, \bibinfo {author} {\bibfnamefont {C.-G.}\
  \bibnamefont {Duan}}, \bibinfo {author} {\bibfnamefont {X.}~\bibnamefont
  {Lu}}, \bibinfo {author} {\bibfnamefont {H.}~\bibnamefont {Luo}}, \bibinfo
  {author} {\bibfnamefont {P.}~\bibnamefont {Dai}}, \ and\ \bibinfo {author}
  {\bibfnamefont {X.-J.}\ \bibnamefont {Chen}},\ }\enquote {\bibinfo {title}
  {Nodeless superconductivity in the presence of spin-density wave in pnictide
  superconductors: The case of
  ${\mathrm{BaFe}}_{2\ensuremath{-}x}{\mathrm{Ni}}_{x}{\mathrm{As}}_{2}$},}\
  \href{\doibase 10.1103/PhysRevB.91.024510} {\bibfield  {journal} {\bibinfo
  {journal} {Phys. Rev. B}\ }\textbf {\bibinfo {volume} {91}},\ \bibinfo
  {pages} {024510} (\bibinfo {year} {2015})}\BibitemShut {NoStop}%
\bibitem [{\citenamefont {Han}\ \emph {et~al.}(2009)\citenamefont {Han},
  \citenamefont {Zhu}, \citenamefont {Cheng}, \citenamefont {Mu}, \citenamefont
  {Jia}, \citenamefont {Fang}, \citenamefont {Wang}, \citenamefont {Luo},
  \citenamefont {Zeng}, \citenamefont {Shen}, \citenamefont {Shan},
  \citenamefont {Ren},\ and\ \citenamefont {Wen}}]{whh}%
  \BibitemOpen
  \bibfield  {author} {\bibinfo {author} {\bibfnamefont {F.}~\bibnamefont
  {Han}}, \bibinfo {author} {\bibfnamefont {X.}~\bibnamefont {Zhu}}, \bibinfo
  {author} {\bibfnamefont {P.}~\bibnamefont {Cheng}}, \bibinfo {author}
  {\bibfnamefont {G.}~\bibnamefont {Mu}}, \bibinfo {author} {\bibfnamefont
  {Y.}~\bibnamefont {Jia}}, \bibinfo {author} {\bibfnamefont {L.}~\bibnamefont
  {Fang}}, \bibinfo {author} {\bibfnamefont {Y.}~\bibnamefont {Wang}}, \bibinfo
  {author} {\bibfnamefont {H.}~\bibnamefont {Luo}}, \bibinfo {author}
  {\bibfnamefont {B.}~\bibnamefont {Zeng}}, \bibinfo {author} {\bibfnamefont
  {B.}~\bibnamefont {Shen}}, \bibinfo {author} {\bibfnamefont {L.}~\bibnamefont
  {Shan}}, \bibinfo {author} {\bibfnamefont {C.}~\bibnamefont {Ren}}, \ and\
  \bibinfo {author} {\bibfnamefont {H.-H.}\ \bibnamefont {Wen}},\ }\enquote
  {\bibinfo {title} {Superconductivity and phase diagrams of the $4d$- and
  $5d$-metal-doped iron arsenides
  ${\text{SrFe}}_{2\ensuremath{-}x}{M}_{x}{\text{As}}_{2}$
  $(M=\text{Rh},\text{Ir},\text{Pd})$},}\ \href{\doibase
  10.1103/PhysRevB.80.024506} {\bibfield  {journal} {\bibinfo  {journal} {Phys.
  Rev. B}\ }\textbf {\bibinfo {volume} {80}},\ \bibinfo {pages} {024506}
  (\bibinfo {year} {2009})}\BibitemShut {NoStop}%
\bibitem [{not()}]{note2}%
  \BibitemOpen
  \href@noop {} {}\bibinfo {note} {For the SFM RuSr$_{2}$GdCu$_{2}$O$_{8}$
  where the SV state is stabilized at the high-temperature side, in contrast,
  the FCC curve has a larger magnetic susceptibility
  \cite{Bernhard}.}\BibitemShut {Stop}%
\bibitem [{\citenamefont {Bean}(1964)}]{bean}%
  \BibitemOpen
  \bibfield  {author} {\bibinfo {author} {\bibfnamefont {C.~P.}\ \bibnamefont
  {Bean}},\ }\enquote {\bibinfo {title} {Magnetization of High-Field
  Superconductors},}\ \href{\doibase 10.1103/RevModPhys.36.31} {\bibfield
  {journal} {\bibinfo  {journal} {Rev. Mod. Phys.}\ }\textbf {\bibinfo {volume}
  {36}},\ \bibinfo {pages} {31} (\bibinfo {year} {1964})}\BibitemShut {NoStop}%
\bibitem [{\citenamefont {Fulde}\ and\ \citenamefont {Ferrell}(1964)}]{FF}%
  \BibitemOpen
  \bibfield  {author} {\bibinfo {author} {\bibfnamefont {P.}~\bibnamefont
  {Fulde}}\ and\ \bibinfo {author} {\bibfnamefont {R.~A.}\ \bibnamefont
  {Ferrell}},\ }\enquote {\bibinfo {title} {Superconductivity in a Strong
  Spin-Exchange Field},}\ \href{\doibase 10.1103/PhysRev.135.A550} {\bibfield
  {journal} {\bibinfo  {journal} {Phys. Rev.}\ }\textbf {\bibinfo {volume}
  {135}},\ \bibinfo {pages} {A550} (\bibinfo {year} {1964})}\BibitemShut
  {NoStop}%
\bibitem [{\citenamefont {Larkin}\ and\ \citenamefont
  {Ovchinnikov}(1965)}]{LO}%
  \BibitemOpen
  \bibfield  {author} {\bibinfo {author} {\bibfnamefont {A.~I.}\ \bibnamefont
  {Larkin}}\ and\ \bibinfo {author} {\bibfnamefont {Y.~N.}\ \bibnamefont
  {Ovchinnikov}},\ }\enquote {\bibinfo {title} {Inhomogeneous State of
  Superconductors},}\ \href{<Go to ISI>://WOS:A19656224900049} {\bibfield
  {journal} {\bibinfo  {journal} {Soviet Physics JETP-USSR}\ }\textbf {\bibinfo
  {volume} {20}},\ \bibinfo {pages} {762} (\bibinfo {year} {1965})}\BibitemShut
  {NoStop}%
\bibitem [{\citenamefont {Liu}\ \emph {et~al.}(2016{\natexlab{a}})\citenamefont
  {Liu}, \citenamefont {Liu}, \citenamefont {Tang}, \citenamefont {Jiang},
  \citenamefont {Wang}, \citenamefont {Ablimit}, \citenamefont {Jiao},
  \citenamefont {Tao}, \citenamefont {Feng}, \citenamefont {Xu},\ and\
  \citenamefont {Cao}}]{Rb1144}%
  \BibitemOpen
  \bibfield  {author} {\bibinfo {author} {\bibfnamefont {Y.}~\bibnamefont
  {Liu}}, \bibinfo {author} {\bibfnamefont {Y.-B.}\ \bibnamefont {Liu}},
  \bibinfo {author} {\bibfnamefont {Z.-T.}\ \bibnamefont {Tang}}, \bibinfo
  {author} {\bibfnamefont {H.}~\bibnamefont {Jiang}}, \bibinfo {author}
  {\bibfnamefont {Z.-C.}\ \bibnamefont {Wang}}, \bibinfo {author}
  {\bibfnamefont {A.}~\bibnamefont {Ablimit}}, \bibinfo {author} {\bibfnamefont
  {W.-H.}\ \bibnamefont {Jiao}}, \bibinfo {author} {\bibfnamefont
  {Q.}~\bibnamefont {Tao}}, \bibinfo {author} {\bibfnamefont {C.-M.}\
  \bibnamefont {Feng}}, \bibinfo {author} {\bibfnamefont {Z.-A.}\ \bibnamefont
  {Xu}}, \ and\ \bibinfo {author} {\bibfnamefont {G.-H.}\ \bibnamefont {Cao}},\
  }\enquote {\bibinfo {title} {Superconductivity and ferromagnetism in
  hole-doped ${\mathrm{RbEuFe}}_{4}{\mathrm{As}}_{4}$},}\ \href{\doibase
  10.1103/PhysRevB.93.214503} {\bibfield  {journal} {\bibinfo  {journal} {Phys.
  Rev. B}\ }\textbf {\bibinfo {volume} {93}},\ \bibinfo {pages} {214503}
  (\bibinfo {year} {2016}{\natexlab{a}})}\BibitemShut {NoStop}%
\bibitem [{\citenamefont {Liu}\ \emph {et~al.}(2016{\natexlab{b}})\citenamefont
  {Liu}, \citenamefont {Liu}, \citenamefont {Chen}, \citenamefont {Tang},
  \citenamefont {Jiao}, \citenamefont {Tao}, \citenamefont {Xu},\ and\
  \citenamefont {Cao}}]{Cs1144}%
  \BibitemOpen
  \bibfield  {author} {\bibinfo {author} {\bibfnamefont {Y.}~\bibnamefont
  {Liu}}, \bibinfo {author} {\bibfnamefont {Y.-B.}\ \bibnamefont {Liu}},
  \bibinfo {author} {\bibfnamefont {Q.}~\bibnamefont {Chen}}, \bibinfo {author}
  {\bibfnamefont {Z.-T.}\ \bibnamefont {Tang}}, \bibinfo {author}
  {\bibfnamefont {W.-H.}\ \bibnamefont {Jiao}}, \bibinfo {author}
  {\bibfnamefont {Q.}~\bibnamefont {Tao}}, \bibinfo {author} {\bibfnamefont
  {Z.-A.}\ \bibnamefont {Xu}}, \ and\ \bibinfo {author} {\bibfnamefont {G.-H.}\
  \bibnamefont {Cao}},\ }\enquote {\bibinfo {title} {A new ferromagnetic
  superconductor: {CsEuFe$_4$As$_4$}},}\ \href{\doibase
  http://dx.doi.org/10.1007/s11434-016-1139-2} {\bibfield  {journal} {\bibinfo
  {journal} {Sci. Bull.}\ }\textbf {\bibinfo {volume} {61}},\ \bibinfo {pages}
  {1213 } (\bibinfo {year} {2016}{\natexlab{b}})}\BibitemShut {NoStop}%
\bibitem [{\citenamefont {Meier}\ \emph {et~al.}(2016)\citenamefont {Meier},
  \citenamefont {Kong}, \citenamefont {Kaluarachchi}, \citenamefont {Taufour},
  \citenamefont {Jo}, \citenamefont {Drachuck}, \citenamefont {B\"ohmer},
  \citenamefont {Saunders}, \citenamefont {Sapkota}, \citenamefont {Kreyssig},
  \citenamefont {Tanatar}, \citenamefont {Prozorov}, \citenamefont {Goldman},
  \citenamefont {Balakirev}, \citenamefont {Gurevich}, \citenamefont {Bud'ko},\
  and\ \citenamefont {Canfield}}]{canfield}%
  \BibitemOpen
  \bibfield  {author} {\bibinfo {author} {\bibfnamefont {W.~R.}\ \bibnamefont
  {Meier}}, \bibinfo {author} {\bibfnamefont {T.}~\bibnamefont {Kong}},
  \bibinfo {author} {\bibfnamefont {U.~S.}\ \bibnamefont {Kaluarachchi}},
  \bibinfo {author} {\bibfnamefont {V.}~\bibnamefont {Taufour}}, \bibinfo
  {author} {\bibfnamefont {N.~H.}\ \bibnamefont {Jo}}, \bibinfo {author}
  {\bibfnamefont {G.}~\bibnamefont {Drachuck}}, \bibinfo {author}
  {\bibfnamefont {A.~E.}\ \bibnamefont {B\"ohmer}}, \bibinfo {author}
  {\bibfnamefont {S.~M.}\ \bibnamefont {Saunders}}, \bibinfo {author}
  {\bibfnamefont {A.}~\bibnamefont {Sapkota}}, \bibinfo {author} {\bibfnamefont
  {A.}~\bibnamefont {Kreyssig}}, \bibinfo {author} {\bibfnamefont {M.~A.}\
  \bibnamefont {Tanatar}}, \bibinfo {author} {\bibfnamefont {R.}~\bibnamefont
  {Prozorov}}, \bibinfo {author} {\bibfnamefont {A.~I.}\ \bibnamefont
  {Goldman}}, \bibinfo {author} {\bibfnamefont {F.~F.}\ \bibnamefont
  {Balakirev}}, \bibinfo {author} {\bibfnamefont {A.}~\bibnamefont {Gurevich}},
  \bibinfo {author} {\bibfnamefont {S.~L.}\ \bibnamefont {Bud'ko}}, \ and\
  \bibinfo {author} {\bibfnamefont {P.~C.}\ \bibnamefont {Canfield}},\
  }\enquote {\bibinfo {title} {Anisotropic thermodynamic and transport
  properties of single-crystalline ${\mathrm{CaKFe}}_{4}{\mathrm{As}}_{4}$},}\
  \href{\doibase 10.1103/PhysRevB.94.064501} {\bibfield  {journal} {\bibinfo
  {journal} {Phys. Rev. B}\ }\textbf {\bibinfo {volume} {94}},\ \bibinfo
  {pages} {064501} (\bibinfo {year} {2016})}\BibitemShut {NoStop}%
\bibitem [{\citenamefont {Ren}\ \emph {et~al.}(2008)\citenamefont {Ren},
  \citenamefont {Zhu}, \citenamefont {Jiang}, \citenamefont {Xu}, \citenamefont
  {Tao}, \citenamefont {Wang}, \citenamefont {Feng}, \citenamefont {Cao},\ and\
  \citenamefont {Xu}}]{Eu122-rz}%
  \BibitemOpen
  \bibfield  {author} {\bibinfo {author} {\bibfnamefont {Z.}~\bibnamefont
  {Ren}}, \bibinfo {author} {\bibfnamefont {Z.}~\bibnamefont {Zhu}}, \bibinfo
  {author} {\bibfnamefont {S.}~\bibnamefont {Jiang}}, \bibinfo {author}
  {\bibfnamefont {X.}~\bibnamefont {Xu}}, \bibinfo {author} {\bibfnamefont
  {Q.}~\bibnamefont {Tao}}, \bibinfo {author} {\bibfnamefont {C.}~\bibnamefont
  {Wang}}, \bibinfo {author} {\bibfnamefont {C.}~\bibnamefont {Feng}}, \bibinfo
  {author} {\bibfnamefont {G.}~\bibnamefont {Cao}}, \ and\ \bibinfo {author}
  {\bibfnamefont {Z.}~\bibnamefont {Xu}},\ }\enquote {\bibinfo {title}
  {Antiferromagnetic transition in ${\text{EuFe}}_{2}{\text{As}}_{2}$: A
  possible parent compound for superconductors},}\ \href{\doibase
  10.1103/PhysRevB.78.052501} {\bibfield  {journal} {\bibinfo  {journal} {Phys.
  Rev. B}\ }\textbf {\bibinfo {volume} {78}},\ \bibinfo {pages} {052501}
  (\bibinfo {year} {2008})}\BibitemShut {NoStop}%
\end{thebibliography}%
\newpage

\textbf{Acknowledgments}
This work is supported by the National Science Foundation of China (Nos.11474252 and 11504329), the National Key Research and Development Program of China (No.2016YFA0300202), and the Fundamental Research Funds for the Central Universities of China.

\textbf{Author contributions}
W.H.J. and Y.L. grew and characterized the crystals. W.H.J. and Q.T. performed the magnetic measurements. W.H.J., Z.R. and G.H.C. analyzed and interpreted the data, and wrote the paper. This work was coordinated and designed by G.H.C.

\textbf{Additional Information}
The authors declare no competing financial interests.

\end{document}